\documentstyle[titlepage,fleqn,11pt]{article}
\begin{document}
\begin{titlepage}
\title{ On mean free path statistics of relativistic
heavy ions in nuclear emulsion}
\author{Haret C. Rosu \\
\\
Instituto de F\'{\i}sica, Universidad de Guanajuato\\
Apartado Postal E-143, 37150 L\'eon, Gto, M\'exico}
\date{}
{\baselineskip=20pt

\begin{abstract}
We survey at the general level, and in a rather wide and
historical context,
some less known mathematical details of
various mean free path statistics in nuclear emulsion as applied to
relativistic projectile fragmentations.
A number of comments are provided on Feller's paradox, ordered, censored,
and truncated statistics, Erlang and Poisson observers of nuclear emulsion.
All these issues are related to
the statistical approaches used about a decade ago at Berkeley Bevalac for
anomalons,
some of which should be considered as standard methods for mean free paths'
estimators of relativistic heavy ions.

\vskip 2cm


\vskip 2cm

Internet: rosu@ifug.ugto.mx

\vskip 1.5cm

IFUG-10/94; version published in Acta Physica Polonica B 25, 1459
(October 1994)

\vskip 1cm

hep-ph/9412263

\end{abstract}

    }
\maketitle
\end{titlepage}
\baselineskip=30pt
\parskip=0pt

The mean free path $\lambda$ is a statistical concept introduced by
Clausius in 1858 that appears in physics in two situations:

i. In the qualitative evaluation of kinetic phenomena through the
following approximation of the collision term in Boltzmann equation
$$ Coll(f)\approx -(f-f_{0})/\tau\approx -\frac{\bar{v}}{\lambda}(f-f_{0})
  \eqno(1)  $$

ii. In detection problems, where $\lambda$ is the parameter of the
exponential distribution of interacting distances
$$f(x)dx=\exp(-x/\lambda)dx/\lambda  \eqno(2)  $$
In the following we shall be interested in determining the mean free paths
of relativistic heavy ions in nuclear emulsion. The first tracks of
relativistic heavy ions in emulsion stacks were obtained in 1948 in
cosmic ray experiments by Bradt and Peters \cite{bp}. Shortly after,
in the period 1954-1957, well-known authors in cosmic ray physics
reported mean free paths smaller than expected for some isolated long
heavy-ion fragmentation cascades but this was considered as being due
to the highly biased sample of events recorded by a detector of limited
dimensions like an emulsion stack \cite{hw}.

The mean free path and the interaction cross-section are connected by
$\lambda=(n\sigma)^{-1}$ where n is the concentration of the scattering
centres. This simply corresponds to the case of one molecule moving
through the rest of the molecules taken as a fixed background in
molecular physics. Since emulsion is a heterogeneous medium, one should
make use of the more general formula
$\lambda=\sum (n_{i}\sigma_{i})^{-1}$, where $n_{i}$ corresponds to the
composition of the emulsion in use. For $\sigma_{i}$ one can consider
the geometrical cross section from molecular physics with the difference
that the nuclear interaction is present only when a certain overlap of
the cross sections occurs (Bradt-Peters formula)
$$\sigma_{BP}=\pi (R_{p}^{2}+R_{t}^{2}-\delta)=\pi r_{0}^{2}(A_{p}^{1/3}
+A_{t}^{1/3}-b)^{2}  \eqno(3) $$
The indices $p$ and $t$ stand for projectile and target. The overlap
parameter $b$ can be related to the radius parameter $r_{0}$. In nuclear
physics the radius parameter is in the range
$r_{0}=(1.15\div 1.45)\cdot 10^{-13}$ cm. In this way, one can obtain
a theoretical (Bradt-Peters) mean free path and compare it with the
experimental (statistical) one. Geometrical arguments are very powerful
as far as quantum mechanics does not come into play too much,
as it is the case of projectile
fragmentations in nuclear emulsion. However we face the problem of the
precise experimental determination of the mean free path. One can think of
two methods for that:

i. For a big number of events, one can plot the exponential distribution
of the interaction paths and determine the mean free path graphically.
This method is too empirical.

ii. A very general method is certainly based on the maximum likelihood
function in its various forms.

Thirty years ago it was in vogue \cite{da} to use the so-called
Bartlett functions $S$ and $S^{'}$ \cite{bar}
In this procedure, the Bartlett function $S^{'}$ is
plotted against $1/\lambda$. This function is normally distributed with
zero mean and dispersion equal to one. The maximum likelihood value of
the mean free path is given by the equation $S'(\lambda^{*})$, that is
the intersection with the $1/\lambda$ axis. Moreover the errors (one
standard deviation) are given by $S^{'}(\lambda)=\pm 1$. Unfortunately
the expression for $S^{'}(\lambda)$ is very complicated. Some time ago,
Yost \cite{yo} generalized the results of Bartlett to the case when the
observations are affected by experimental errors. These are important
in the time of flight measurements, but for nuclear
emulsions the errors of the measured distances are negligible.

New statistical methods have been introduced in 1980, when
Friedl\"ander et al. \cite{fr} presented their interesting statistical
treatment of Bevalac projectile fragmentations in Ilford G5 emulsions.
Despite their conclusions concerning the {\em anomalons}, the
Berkeley works have been a natural continuation of the cosmic ray works in the
fifties with the essential difference that the emulsion was irradiated
with relativistic heavy ion beams from Berkeley Bevalac. Moreover,
emulsion pellicles were exposed to 200 A GeV $O^{16}$ and $S^{32}$ ion
beams at CERN SPS \cite{pe} and it is to be expected that emulsion
groups will take their part in future heavy ion relativistic experiments.
Some of the statistical methods of Friedl\"ander et al.
should be considered as standard ones for mean free paths estimators
of heavy ions in emulsion at all these  energies.

The first systematic studies of the heavy-ion mean free paths in nuclear
emulsion were done by Judek \cite{ju} who reported anomalies
in the case of projectile fragments
(Judek effect/{\em anomalons}). The method
used by Judek and later by Friedl\"ander et al. to obtain the
mean free path is based on the formula
$$\lambda ^{*}= T/n  \eqno(4)  $$
where $\lambda ^{*}$ is the maximum likelihood estimate, $T$ is the
total length to be watched until one gets $n$ interactions (emulsion
stars) on both interacting and non-interacting tracks. Usually $T$
has been taken in bins of 1 cm starting from the mother stars and in
the first 2-3 cm a distance dependent mean free path for projectile
fragments was reported, which has been fitted by means of
{\em anomalons}.

We pass now to the scope of the present work which is to sketch some
less known mathematical details of the mean free
path statistics of relativistic heavy ions in nuclear emulsion.

The first thing one should realize
from the mathematical point of view is that the propagation of
relativistic heavy ion beams in emulsion is a Poisson stochastic point
process along the direction of propagation \cite{ro}.
More exactly, due to some lost interactions it is a
rarefied or damaged Poisson process \cite{gk}. This is directly sugessted
by the visual aspect of the fragmentation processes,
especially for projectile fragmentations, as they are seen by means of a
microscope. At relativistic energies the ionization loss of heavy ions
is much less than their kinetic energies, and what one may see are rather
long ionization tracks interrupted occasionally by some stars, representing
the nuclear interactions/collisions with the emulsion nuclei. Considering
these stars as random points along the propagation/ionization lines one
naturally arrives at the idea of a random sthochastic process in the
space variable. Since the distances
between points are exponentially distributed, the stochastic process is by
definition a Poisson point process.

We would like first to worn on the presence of paradoxes.
Stochastic processes as the entire theory of probabilities are affected
by paradoxes generated by the intricacies of the statistical way of
thinking. For Poisson processes, Zernike discussed already in 1928
the so-called {\em `Wegl\"angenparadoxon'} \cite{z}. A form of Zernike's
paradox is well-known in solid state physics with regard to the scattering
of electrons in crystals \cite{ssp}. It refers to the mean time interval
between the last and next collision for the Poisson electron process.
This is about twice the normal average time between collisions.
On the other hand in the well-known
book of Feller \cite{fe} one can find the {\em waiting time paradox}.
Previously we applied the {\em waiting time paradox} (or more
appropriately for emulsion - {\em waiting distance paradox})
to the anomalon artifact \cite{ro}. This paradox shows up whenever
the measurements are performed from the very beginning of the Poisson
process, which consequently is not yet stationary (ergodic). It happens
that this is just the case of projectile fragments in emulsion within
the statistical treatment of Judek and later of Friedl\"ander et al.
When this paradox
is taken into account, there comes out that near the origin of the Poisson
process, the mean free path is distance dependent in the following way
$$ \lambda _{w}=\lambda [1-\frac{1}{2}\exp (-x/\lambda)]   \eqno(5)  $$
for $x\ll \lambda$. Hence, whenever one is trying to detect the very first
collision will enter the paradox and will find out a much smaller mean free
path. For the particular case $x=0$, the mean free path will be only half of
the normal one.
Indeed the
Berkeley group noticed that a mean free path twice
smaller than the normal one could be accommodated by the data assuming
that {\em all} projectile fragments are born anomalous and shortly
after disappear. So, this is just the statement of the
{\em waiting distance paradox} \cite{ro} \cite{fe}. Why does not the
paradox show up for the primaries ? In this case one may invoke
the loss of memory property of Poisson processes, and so to say, is
old enough to be stationary. The point is that for
the projectile fragments the `mother stars' play the role of
the origin of the stochastic process, and therefore one should be
cautious with the non-stationarity property. A complication arises due
to the fact that there is not one stochastic process, but several ones
corresponding to the different charge values \cite{ro}.

The most important statistical test used by the Berkeley group,
considered to be independent of parametrization
($\lambda_{Z}\approx \Lambda Z^{-b}$) is an F test in the disguised
form of the integral probability, which is known to be uniformly
distributed between 0 and 1 by a textbook theorem \cite{fe}. However
the F test, as a variance ratio test, will check only the normality
of the maximum likelihood estimates around the true mean free path.
These estimates are not ideal gaussian variables in the case of small
samples like those ones of fixed Z and at small distances from the
origin. Besides, the F test is known not to be a very robust test
against deviations from normality \cite{jl}. This non-robustness of
the F test is clearly demonstrated by employing other tests. For
example, the null hypothesis of no anomalons is not rejected at a
confidence level of 0.95 by a more robust test, like the
Kolmogorov-Smirnov test which is based on the largest deviation
of the empirical distribution from the expected cumulative one
\cite{du}. Moreover the K-S test has also the advantage of being
independent of the theoretical distribution.

Pshenin and Voinov \cite{pv} have shown that for small samples,
the maximum likelihood estimate for $\lambda$ at fixed number of
primary tracks (say N) and variable number of interactions (say n)
is biased to lower values, according to the following formula
$$B=\lambda^{*}-\lambda=-D/[1-\exp(-D/\lambda)]+ND<1/n>  \eqno(6) $$
where B is the bias and D is the censoring distance. This result
has been known since a long time in statistics \cite{ba} \cite{ml}.
The exact
formula for the mean $<1/n>$ is not very simple but one can use the
approximation \cite{gr}
$$<1/n>= [N(1-\exp(-D/\lambda))-\exp(-D/\lambda)]^{-1}  \eqno(7) $$
working very well for values X= $N(1-\exp(-D/\lambda) > 10$.
An even more accurate result in the range X > 5 is given in \cite{ml}
$$<1/n>=\frac{N-2}{N}[(N-1)(1-\exp(-D/\lambda))]^{-1}  \eqno(8) $$
All these results do apply only for very small samples.

Garpman et al. \cite{ga} proposed as the maximum likelihood estimate
$$\lambda^{*}= T/(n+1)   \eqno(9) $$
The reason is that this estimate is finite also for n=0 and the Monte
Carlo simulation is better. This is just like saying that if you did not
find any interaction the mean free path is equal to the measured
distance.

We now go on with a short presentation of censored and truncated
distributions of the emulsion tracks. The censored distribution is
the distribution of both interacting and noninteracting tracks, while
the truncated one is based only on the interacting tracks included
in the censoring/truncating distance D. One might think that the
censoring distribution carries more information. However this is
not so. Both the censored and truncated distributions lead to the
same maximum likelihood estimate. This can be shown by means of
results from the ordered statistics \cite{jl}.

Suppose there are k interacting tracks from a total of N tracks within
the truncation distance and we order them from smaller to larger ones:
$0<x_{1}<x_{2}<.....<x_{k}<D$. Then the maximum likelihood function
constructed from these distances only is \cite{jl}
$$L=\frac{n!}{(n-k)!}\lambda^{-k}\exp(-y_{k}/\lambda)  \eqno(10) $$
where
$$y_{k}=\sum _{j=1}^{k-1} x_{j} +(n-k+1)x_{k}  \eqno(11) $$
The resulting ordered maximum likelihood estimate will be
$$\lambda_{o}^{*}=k^{-1}\sum_{j=1}^{k-1}x_{j}+ k^{-1}(n-k-1)x_{k}=
k^{-1}\sum_{j=1}^{k}x_{j}+k^{-1}(n-k)x_{k}   \eqno(12)  $$
This estimate is practically the same as the censored one which is
given by
$$\lambda_{c}^{*}=k^{-1}\sum_{j=1}^{k}x_{j}+k^{-1}(n-k)D  \eqno(13) $$
for the same sample of k events. Actually censoring can be done in
a number of ways \cite{cl}. In censoring of type I, the censoring
distance is fixed and the number of events is a random variable. In
censoring of type II, the number of events k is fixed and the
censoring distances $x_{k}$ become random variables. The ordered
estimate can be considered as a censored one of type II, whereas the
Berkeley estimate is of type I. A third type of censoring, in which
the censoring distance is different from track to track (so-called
random censoring used in life test data \cite{co}) gives no new result
for the estimation problem. There are only minor differences in the
estimate variances for all types of censoring \cite{cl}.

Finally, we want to comment on the type of observers paying attention
to a stochastic process. This problem is discussed by Egon \cite{eg}
in connection with the theory of signal detection. In the case of
emulsion detectors, the common observers are gamma (Erlang) observers.
They scan the Poisson process until a certain number of events are
registered. For them the distances between the events are important in
doing their statistics. Poisson observers instead are looking for a
preassigned number of events in a given distance interval. However
the estimation problem gives the same result as for
the Erlang observers. This is so because in both cases actually what
we have is the same family of exponential distributions.

\section*{Acknowledgments}

This work was partially supported by CONACyT Grant No. F246-E9207 to the
University of Guanajuato. The author thanks to the referee for an important
remark.

\end{document}